\documentclass[usenatbib]{mn2e}
\usepackage{graphicx}
\usepackage{subfigure}
\bibpunct{(}{)}{;}{a}{}{,}
\usepackage{amsmath,amssymb}

\title[Impact of GWB on pulsar timing parameters]{The impact of a stochastic gravitational-wave background on pulsar timing parameters}
\author[J. A. Ellis et. al.]{J. A. Ellis$^{\dagger1}$, M. A. McLaughlin$^{1,2,3}$ and J.~P.~W. Verbiest$^{4}$\\
$^{\dagger}$Inquiries to: justin.ellis18@gmail.com\\
$^{1}$Department of Physics, West Virginia University, Morgantown, WV 26506, USA.\\
$^2$Also adjunct at the National Radio Astronomy Observatory, Green Bank, WV 24944, USA. \\ 
$^3$Alfred P. Sloan Research Fellow. \\
$^{4}$Max-Planck-Institut f\"{u}r Radioastronomie, Auf dem H\"{u}gel 69, 53121 Bonn, Germany.}
\begin{document}


\pagerange{000--000} \pubyear{0000}

\maketitle

\label{firstpage}

\begin{abstract}
Gravitational waves are predicted by Einstein's theory  of general relativity as well as other theories of gravity. The rotational stability of the fastest pulsars means that timing of an array of these objects can be used to detect and investigate gravitational waves. Simultaneously, however, pulsar timing is used to estimate spin period, period derivative, astrometric, and binary parameters. Here we calculate the effects that a stochastic background of gravitational waves has on pulsar timing parameters through the use of simulations { and}{ data from the millisecond pulsars PSR J0437--4715 and PSR J1713+0747.} We show that the reported timing uncertainties become underestimated with increasing background amplitude { by up to a factor of $\sim10$ for a stochastic gravitational-wave background amplitude of $A=5\times 10^{-15}$, { where $A$ is the amplitude of the characteristic strain spectrum at one-year gravitational wave periods}}. { We find evidence for  prominent low-frequency spectral leakage in simulated data sets including a stochastic gravitational-wave background.} We use these simulations along with independent Very Long Baseline Interferometry (VLBI) measurements of parallax to set a 2--sigma upper limit of $A\le9.1\times 10^{-14}$.  We find that different supermassive black hole assembly scenarios do not have a significant effect on the calculated upper limits.  We also test the effects that ultralow--frequency (10$^{-12}$--10$^{-9}$ Hz) gravitational waves have on binary pulsar parameter measurements and find that the corruption of these parameters is less than { those due to $10^{-9}$--$10^{-7}$ Hz gravitational waves.}
\end{abstract}

\section{Introduction}
\label{intro}
Pulsars are highly magnetized, rapidly rotating neutron stars that have spin periods ranging from  seconds to milliseconds. See \citet{handbook} for a full review of pulsars and their applications. Pulsars can be separated into two broad categories: normal pulsars and millisecond pulsars (MSPs). Normal pulsars are characterized by large periods and period derivatives ($P\sim 1$ s and $\dot{P}\sim 10^{-12}$, respectively) and large magnetic fields ($B\sim 10^{12}$ G), whereas MSPs have smaller periods and period derivatives ($P\sim 0.01$ s and $\dot{P}\sim 10^{-20}$, respectively) and smaller magnetic fields ($B\sim 10^{8}$ G). Pulsars are extremely accurate clocks and, specifically, MSPs exhibit much better timing stability as they do not show evidence of rotational instabilities such as timing noise or glitches, which are prevalent in normal pulsars. In fact, many MSPs rival atomic clocks in their fractional stability (see Figure 24 of \citealt{l08}).  Because of this and the fact that many MSPs are in binary systems, they are extensively used for tests of general relativity. The first evidence for the existence of  gravitational waves (GWs) come from the precise timing of PSR B1913+16 in which the decrease of  the orbital period is in accordance with energy loss due to gravitational radiation, as predicted by general relativity \citep{tw82}. 

Nearly three decades ago, it was first shown that pulsars could be used to directly detect GWs \citep{saz78,det79} by using the pulsar timing residuals (further discussed in Section \ref{sim}) to look for a specific GW signature. \citet{hd83} showed that by exploiting the known correlations that a stochastic GW background (GWB) would induce in the pulsar timing residuals, one can place an upper limit on the GWB. This work introduces what is today known as the Hellings and Downs curve, which is a main feature in many detection algorithms (see, e.g., \citealt{jhl+05,abc+08}). The concept of a pulsar timing array (PTA) composed of the best timed MSPs was first developed over two decades ago \citep{r89,fb90}. Today there are three main PTAs in existence with the goal of GW detection using pulsars: the European Pulsar Timing Array (EPTA; \citealt{jsk+08}), the North American Nanohertz Observatory for Gravitational waves (NANOGrav; \citealt{jfl+09}), and the Parkes Pulsar Timing Array (PPTA; \citealt{m08}), all of which are in collaboration to form the International Pulsar Timing Array (IPTA; \citealt{haa+10}).

\subsection{Pulsar timing precision and parameter uncertainties}
While the high timing precision of MSPs has allowed the precise measurement of many parameters and effects, pulsar timing suffers from the inherent fact that the noise sources that contribute to the timing residuals are not well understood. To first order, our timing residuals are composed of noise that follows Gaussian statistics and that has a white power spectrum. For each time-of-arrival (TOA), the amount of receiver noise is estimated by the TOA uncertainty, as determined from a least-squares fit to a high signal-to-noise ratio template profile. It is, however, common for these uncertainties to be underestimated or for additional sources of white noise to contribute to the timing residuals.  One reason might be self-standarding: in the process of deriving an average TOA for an observation, the integrated pulse profile is correlated with a template profile. If the template profile is generated from addition of many observations, then at some low level the noise from the observation can correlate with the noise in the standard, leading to an underestimation of the TOA uncertainty. Also, the cross-correlation method typically used has been shown to underestimate TOA uncertainties in the high-noise regime \citep{hbo05}. Poor selection of standard profiles can underestimate the TOA uncertainty as well. All of these effects cause the pulsar timing parameters to be consequently more uncertain than initially expected. Assuming these effects scale with the uncertainties they affect, one can attempt to mitigate this underestimation by multiplication of the TOA uncertainties by the square root of the reduced $\chi^2$ value of the timing residuals, as is commonly done in other fields.

Underestimation of the white noise contribution is not the only challenge in assessing pulsar timing parameters and their true uncertainties, though. The presence of non-white noise in pulsar timing residuals is very common for young pulsars and, as timing baselines increase, starts to become important in timing of some MSPs as well \citep{sc10,vbs+08,sns+05}. While the spectral properties and origins of non-white noise in MSPs are as yet not understood because the white noise is still too prominent to allow detailed analysis, it has already been shown that the least-squares fitting performed in pulsar timing reports underestimated uncertainties for the parameters in the timing fit and may even corrupt the parameter values themselves \citep{vbs+08}.  A technique to improve the least-squares fitting process to mitigate such effects has recently been proposed \citep{chc+10}.

A third problem may be present in the form of signals that are part of a non-white spectrum -- such as GWs. Not only does this type of signal affect the fitting process as described above, through introduction of power at a very specific frequency it has the potential to be fully absorbed in the timing signature of periodic parameters in the pulsar timing model, such as pulsar position, proper motion, parallax or many binary parameters. On the one hand this can prove useful since independent measurement of parameters such as parallax (by, e.g., VLBI) can be combined with the pulsar timing measurement to place bounds on the amount of corrupting noise present (see, e.g. \citealt{dvt+08}), but on the other hand it is worthwhile to assess the amount of influence such noise (especially predicted noise sources such as the GWB) can have on pulsar timing parameters.

\subsection{Pulsar timing and GWs}
\label{sec:ptgw}
 Low frequency ($10^{-9}$--$10^{-7}$ Hz) GWs are expected from supermassive black hole binary systems (SMBHBs), cosmic strings, and GWs from the big bang and inflationary era of the early universe. GWs from these sources can manifest themselves in different ways. Single nearby SMBHBs can produce  resolvable waves  with periods on the order of years \citep{wl03}. SMBHBs and cosmic strings can also produce GW bursts \citep{dv01,smc07,lss09} in which the duration of the GW signal is much less than the observation time. We also expect pulsar timing arrays to be sensitive to a stochastic background of unresolvable sources. This stochastic background can be described by a characteristic strain spectrum $h_c(f)$ defined as a frequency--dependent power law \citep{jhs+06}
\begin{equation}
\label{eq:strain}
h_c(f)=A\left(\frac{f}{\hbox{yr}^{-1}}\right)^{\alpha}.
\end{equation}
Here $A$ is the dimensionless amplitude and $\alpha$ is the spectral index. The power in the GWB can then subsequently be written as
\begin{equation}
\label{eq:power}
P(f)=\frac{1}{12\pi^2}\frac{1}{f^3}h_c(f)^2.
\end{equation}
In accordance with many cosmological theories, one can also write $\Omega_{{\rm gw}}(f)$, the energy density per unit logarithmic frequency interval, in terms of the characteristic strain spectrum as 
\begin{equation}
\label{eq:omega}
\Omega_{{\rm gw}}(f)=\frac{2}{3}\frac{\pi^2}{H_0^2}f^2h_c(f)^2,
\end{equation}
where $H_0$ is the Hubble constant. This paper will  focus primarily on a stochastic background of SMBHB sources with a spectral index of  $\alpha = -2/3$ \citep{wl03,jb03,ein+04}. Recent work has shown that this simple power law does not hold true at higher frequencies ($f> 10^{-8}$) due to the discrete number of sources. At these frequencies the characteristic strain is written as \citep{svc08}
\begin{equation}
\label{eq:broken}
h_c(f)=h_0\left(\frac{f}{f_0}\right)^{\alpha}\left(1+\frac{f}{f_0}\right)^{\gamma},
\end{equation}
where $h_0$, $f_0$, and $\gamma$ are model-dependent parameters based on specific SMBHB merger scenarios. The implications that this result will have on our simulations will be discussed  in Section \ref{upper} where we test whether this broken power law will make a significant difference to methods that use periodic pulsar parameters with relatively high-frequency fitting functions.

Although a direct detection of the GWB has not been made to date, methods using  pulsar timing residuals have been used to disprove proposed SMBHB systems \citep{jll+04} and to put limits on SMBHB coalescence rates \citep{wjy+11} and on the dimensionless strain amplitude \citep{srt+90,jhs+06,hlj+11}. The most stringent published upper limit to date is  $A\le 1.1\times 10^{-14}$  \citep{jhs+06}.  They use a statistic sensitive to a red spectrum.  In this paper, we present a method which relies on the derived timing \citep{vbc+09} and VLBI \citep{dvt+08} parameters  for individual pulsars. It involves adding simulated GWs to pulsar timing parameters and refitting for the pulsar timing parameters to determine the effect of the induced GW.  This idea of using VLBI and timing parameters to place limits on the stochastic background was first introduced in \citet{dvt+08}. In that work, the kinematic distance obtained from timing measurements of $\dot{P}_{\rm b}$ in the presence of a GWB is compared to the VLBI measurement of the distance to place an upper limit.

\subsection{This paper}
\label{sec:paper}
In this paper, we determine how strongly  pulsar parameters can be corrupted by a GWB and red noise in general. We also use independent VLBI measurements to construct upper limits on the stochastic GWB. This is an important question because these corruptions shown here through simulations could be confirmed  by more accurate VLBI measurements as well as other optical interferometry missions such as GAIA \citep{mk10}. Furthermore, many relativistic tests in the strong field regime could be affected by using the corrupted values of the post-Keplerian parameters $\dot{P}_{\rm b}$ and $\dot{\omega}$.

This paper is organized as follows: in Section \ref{sim} we describe the basis of our simulations, including a brief review of GW effects on the pulsar timing residuals and fitting procedures. In Section \ref{obs} we describe the observations used in our simulations. In Section \ref{sec:analysis} we present the effect of GWs on pulsar timing parameters and in Section \ref{discussion} we summarise our findings.

\section{Simulation}
\label{sim}
All simulations and timing analyses for this paper were performed using the \textsc{Tempo2} software package \citep{hem06}. Full details of the timing model and algorithms employed by \textsc{Tempo2} can be found in \citet{ehm06}. 

\subsection{Pulsar timing theory}
Here we will only briefly review the timing model and the standard fitting procedure. TOAs for a specific pulsar  are produced at each epoch by a cross-correlation of the pulse profile with a low noise standard template. This procedure produces site-arrival-times (SATs), which are the times that the average pulse reached the telescope. The SATs are then converted to barycentric-arrival-times (BATs) using the Solar System Ephemeris DE405 \citep{sta04} which gives accurate predictions of the locations and masses of Solar System objects and the position of the telescope at any point in time. The timing model used in \textsc{Tempo2} requires calculating the time of emission at the pulsar
\begin{equation}
t_{\rm e}^{\rm psr}=t_{\rm a}^{\rm obs}-\Delta_{\odot}-\Delta_{\rm IS}-\Delta_{\rm B},
\end{equation}
where $t_{\rm a}^{\rm obs}-\Delta_{\odot}=t_{\rm a}^{\rm SSB}$ is the BAT, $\Delta_{\odot}$ includes all terms that go into converting the observed TOA to the solar system barycenter (SSB), (Roemer, Einstein, and Shapiro delays etc.), $\Delta_{\rm IS}$ includes delays due to passage through the interstellar medium, and $\Delta_{\rm B}$ includes delays related to the pulsar's binary motion. Following the notation of \citet{ehm06}, the subscript ``$a$'' represents an arrival time and the ``$e$'' represents an emission time.
 In order to calculate the timing residuals, the pulse phase $\phi(t)$ must be calculated and compared to the nearest integer value. The function $\phi(t)$ describes the time evolution of the pulse phase and is written in terms of a power series in time
\begin{equation}
\phi(t)=\phi_0+\nu\Delta t+\frac{1}{2}\dot{\nu}\Delta t^2+\frac{1}{6}\ddot{\nu}\Delta t^3+...,
\end{equation}
where $\Delta t=t_{\rm e}^{\rm psr}-t_0$ is the difference in the time of emission from the pulsar and some reference epoch. From this it can be seen that fitting for the pulsar frequency $\nu$ removes a linear offset from the timing residuals and fitting for the frequency derivative $\dot{\nu}$ removes a quadratic function. Although $\phi(t)$ is expressed as a power series in $\Delta t$, second order approximations are very good for MSPs since they have negligible $\ddot{\nu}$.  The fractional part of $\phi(t)$ is the ``residual''. The fitting procedure for this is a $\chi^2$ minimisation where
\begin{equation}
\chi^2=\frac{1}{\nu^{2}}\sum_i\left(\frac{\phi(t_i)-N_i}{\sigma_i}\right)^2.
\end{equation}
Here $N_i$ is the nearest integer to $\phi(t_i)$, $\sigma_i$ is the TOA uncertainty, and $\nu$ is the pulse frequency.  Since the fitting procedure is a $\chi^2$ minimisation technique, the \textsc{Tempo2}-reported uncertainty is just the corresponding statistical uncertainty on the reduced $\chi^2$ fit for any given parameter \citep{pvt+92}.

\subsection{Basic GW simulations}
Here we will give a brief review of the method through which \textsc{Tempo2} adds a stochastic GWB to the pulsar timing residuals (see \citet{hjl+09} for full details). \textsc{Tempo2} uses a time domain method to inject timing residuals caused by GWs. A stochastic background of GWs can be written as a sum of plane waves originating from random positions on the sky. The metric perturbation can then be written as
\begin{equation} 
h_{\mu \nu}={\rm Re}\left[\sum_{j=0}^{N-1}A_{\mu \nu j}{\rm e}^{i(\mathbf{k}_j\cdot \mathbf{x}-\omega_j t)}\right],
\end{equation}
where $N$ is the number of GW sources and $\mathbf{x}$, $t$, $A_{\mu \nu}$, $\mathbf{k}_j$, and $\omega_j$ are the position vector of the source, time,  amplitude, wave number, and angular frequency of the $j$th GW source, respectively. It then follows that the residual induced by this sum of plane GWs is the fractional change in the frequency of the pulse rate integrated over time
\begin{equation}
\begin{split}
R(t)=&-\frac{1}{2}{\rm Re}\Bigg\{\sum_{j=0}^{N-1}i\frac{\hat{k}_p^l\hat{k}_p^mA_{lmj}}{\omega_j}({\rm e}^{-i\omega_j t}-1)\\
&\times\left[\frac{1-{\rm e}^{i\omega_j D(1-\cos\theta_j)}}{1-\cos\theta_j}\right]\Bigg\}.
\end{split}
\end{equation}
Here repeated upper and lower indices indicate a sum over spatial coordinates, $\hat{k}_p$ is the unit vector in the direction of the pulsar, $D$ is the distance from the pulsar to the Earth, $\theta_j$ is the angle between the direction to the pulsar and the $j$th GW source, and $A_{lmj}$ is the transverse-traceless complex amplitude that takes the two GW polarisation states into account. 

\subsection{Extended GW simulations}
\label{sec:extended}
The basic algorithm used in this paper involves creating 1,000 sets of induced timing residuals for each of 200 GWB amplitudes for each pulsar, adding them to the observed residuals and refitting {  for the full timing model \citep{vbs+08,vbc+09}.} Each amplitude is due to an ensemble average of 1,000 individual sources distributed isotropically on the sky. These sources are drawn from a distribution that follows Equation \ref{eq:strain}. The end result of this process is a distribution of the fitted parameters at 200 GWB amplitudes distributed logarithmically in the range $1\times 10^{-15}$--$1\times 10^{-13}$. This range was chosen because it includes all previous upper limits and allows for lower backgrounds. These distributions were then analysed in several ways to calculate the effects of fitting for pulsar parameters when a GW signal is present in the timing residuals and to determine upper limits on the GWB amplitude. It  is likely that there already is  a GW signal from the stochastic background present in the pulsar timing data. Because we could be adding simulated GWs to those already existing in the data, any upper limits that we set could be, at worst, underestimated by a factor of $\sqrt{2}$.

\section{Observations}
\label{obs}
We used data for the pulsars PSR J0437--4715 and PSR J1713+0747. These data were collected with the Parkes 64-m radio telescope at 20 cm wavelength. A full description of the data sets and observing systems can be found in \citet{vbs+08,vbc+09} and references therein. These pulsars were chosen for three reasons. Firstly, the timing parallax measurements are consistent at the 1-$\sigma$ level with those from VLBI, giving us the ability to place upper limits on the background by comparing the timing and VLBI measurements. Secondly, both pulsars have low rms timing residuals, which lead to the best upper limits that can be obtained using the methods described above. Finally, these data sets have a relatively long time span, allowing us to see  effects of low-frequency GWs more readily in the timing residuals. A summary of the timing characteristics is given in Tables \ref{tab:0437} and \ref{tab:1713}. For the VLBI parallax measurements  in this study we use an LBA parallax for PSR J0437--4715 \citep{dvt+08} and a VLBA parallax for PSR J1713+0747 \citep{cbv+09}.

\section{Analysis}
\label{sec:analysis}

\subsection{Absorption of GW power in timing parameters}
\label{errors}
Figure \ref{fig:fits} illustrates how a GWB can be absorbed into certain periodic timing parameters for PSR J0437--4715.  Figure \ref{fig:gw} shows the timing residuals with a GWB with amplitude of $A=5\times 10^{-13}$ added to the data. Figures \ref{fig:pmpx} and \ref{fig:pbom} show the timing residuals using the GW-induced ($A=5\times10^{-13}$) values of parallax and  proper motion, and $\dot{P}_{\rm b}$, $\omega$, and $\dot{\omega}$, respectively. A very large GWB amplitude was used for illustration purposes, but this same effect takes place at smaller amplitudes and is just not as visible by eye.  We can see from the figures that the GWB can induce both high and low-frequency components in residuals. The problem with fitting for these parameters is that, in effect, one is fitting out a significant amount of the GW signal that is contributing to the residuals. { It is important to note that although a stochastic GWB can induce large uncertainties in timing model parameters that are sinusoidally varying at yearly and half-yearly frequencies, a significant amount of the GW power absorbed comes from the lowest frequencies in the data due to spectral leakage. We have subsequently conducted additional simulations in which we only include GWs in a very narrow frequency range around yearly and half-yearly frequencies. In these simulations we see that the corruption to the sinusoidally varying parameters is negligible even for a very large background amplitude of $A=5\times 10^{-13}$. This leads to the conclusion that the absorbed power in the earlier simulations is indeed due to low-frequency spectral leakage and not absorption of power at these higher frequencies. Nevertheless, absorption of high-frequency GW power still occurs, however, the steepness of the GWB power spectrum implies that for any realistic GWB this effect is fully covered up by the radiometer (white) noise.}

\subsection{Impact on timing model parameters}
\label{sec:impact}
 In this section we will construct confidence intervals for the fitted parameters at various GWB amplitudes and show that these confidence intervals become larger as the GWB amplitude increases. The method for calculating these confidence intervals is quite straightforward. The main simulation discussed in Section \ref{sec:extended} is used to obtain distributions of the given fitted parameters at each amplitude. As we have seen above, a GWB will cause excess { low-frequency} power to be absorbed into timing parameters. Examples of the subsequent corruption of the parameters  can be seen in the histograms of Figure \ref{fig:gauss}. 
Since this simulation injects GWs with random sky positions and random frequencies (within a certain range and with a given spectrum), the subsequent corruption of the fitted parameters caused by absorbing parts of the GW spectrum will result in a Gaussian distribution centered around the unperturbed parameter values.  It can be seen from Figure \ref{fig:gauss} that the FWHM of the distribution and therefore the confidence interval on the parameter increases with increasing GWB amplitude. This increased spread in potential parameter estimates is the GW-induced corruption that we aim to quantify in this paper. It must be noted that at sufficiently high GWB amplitudes prominent low-frequency power will be visible in the timing residuals. To ensure phase-connection in our timing, we limited our simulations to a maximum GWB amplitude of $1\times 10^{-13}$. { We have also used phase tracking to take the phase--wraps caused by the injected GWB into account in our simulations.} We also note that towards the higher end of the simulated GWB amplitude range, sufficient levels of low-frequency noise will be present in the simulations to make the uncertainties returned from the standard least-squares fit unreliable \citep[see e.g.][]{vbs+08}. However, since our analysis only uses the best-fit and not its uncertainty, this has no effect on our results.

These histograms can be used to directly study the effects that the presence of GWs has on the parameter estimates resulting from the fit. This is done by plotting the ratio of the standard deviation of the fitted parameters and the \textsc{Tempo2}-reported unperturbed error on the parameters against GWB amplitude.  { The full list of fitted timing parameters along with their values and uncertainties are shown in Tables \ref{tab:0437} and \ref{tab:1713} for the unperturbed case \citep{vbs+08,vbc+09} and the case where a GWB with amplitude at the current upper limit of $A\sim 1\times10^{-14}$ was added to the data.} The results are plotted in Figure \ref{fig:err}. { Firstly, it should be noted that the fits to spin frequency and frequency derivative were by far the most affected by the GWB as can be seen in Tables \ref{tab:0437} and \ref{tab:1713}. However, these parameters are not plotted here as this is expected since the spin frequency and frequency derivative fit out a linear trend and a quadratic trend respectively, thereby absorbing the lowest frequency (highest power) part of the GWB spectrum.  We see that the orbital period derivative, $\dot{P}_{\rm b}$, is affected more for PSR J1713+0747 than for PSR J0437--4715. This is likely due to the difference of orbital periods for these pulsars (5.7 and 67.8 days for PSR J0437--4715 and PSR J1713+0747, respectively). The larger orbital period for PSR J1713+0747  allows for lower frequency power to be absorbed resulting in a larger overall effect by Equation \ref{eq:power}. 

{ Given that the parallax timing signature scales as $D^{-1}\cos{\beta}$, where $\beta$ is the ecliptic latitude and $D$ is the pulsar distance, one would expect the effect of a GWB on the parallax value will scale with the same factor. Using this scaling relation and the known ecliptic latitudes of the pulsars ($30.87^{\circ}$ for PSR J1713+0747 and $-67.87^{\circ}$ for PSR J0437--4715), we expect that the effect of the GWB on the parallax  should be a factor of $\sim3$ times larger for PSR J0437--4715 than for PSR J1713+0747. However, from Tables \ref{tab:0437} and \ref{tab:1713} we see that this ratio is 4.7. We also expect the effect of the GWB on the astrometric position to be stronger for PSR J0437--4715 than for J1713+0747 because of the lower ecliptic of the former. However, we expect this difference to be less than that seen in our simulations. To find the origin of these discrepancies we created simulated data sets for the two pulsars with uniform sampling and bi-monthly observations over 15 years. We then ran our GWB simulation on these data sets and found that the parallax corruption is 3 times worse for PSR J0437--4715 than for PSR J1713+0747 and the astrometric position is slightly more affected for the former, as expected. The reason that the parallax and astrometric position are not so heavily affected in the real data is because of the irregular sampling, which leads to correlations in the fitted parameters. To confirm this, we also simulated data sets with irregular spacing and compared their covariance matrices with those of the uniformly sampled data. We found that the correlations between parameters are lower for the evenly spaced data. }  Finally, it is also important to note that, for both pulsars, the increase in uncertainty of these parameters is approximately linear in GWB amplitude over the region of interest of $1\times 10^{-15}\le A \le 5\times 10^{-14}$.


\subsection{A limit on the GWB strength}
\label{upper}
As discussed in Section \ref{intro}, many papers have been published dealing with calculation of upper limits on the stochastic GWB amplitude. While some calculate this upper limit using a method that is designed to detect the background \citep{vlm+09,hlj+11}, others use a statistic that is exclusively designed  for putting limits on the background \citep{jhs+06}. For example, Jenet et al. use a statistic that is sensitive to a red power spectrum. By design, this method relies on the noise having a very white power spectrum and thus any deviation from Gaussian noise is a problem for this test. Since some non-Gaussian noise may occur in timing data, especially in long data sets, this method is limited in applicability.  However,  it has thus far produced the most stringent published limits. 

The method discussed in this paper is also not a suitable candidate for detection but it does produce independent upper limits that are consistent with previously published limits. Before discussing specific methods, Figure \ref{fig:detection} shows some general results of this method. To obtain these plots, we run the simulation as described in Section \ref{sec:extended} and then take the output at each GWB amplitude and make a histogram as shown in Figure \ref{fig:gauss}. We calculate an upper limit using this scheme by finding the amplitude for which 95\% of the simulations yield parameter values that lie outside of the formal 2--$\sigma$ errors on that parameter. Note here again that the proper motion is significantly affected for both pulsars.  {  However, the orbital period derivative and astrometric positions are strongly affected for PSR J1713+0747 and  PSR J0437--4715 respectively, for reasons discussed in Section \ref{sec:impact}}. While this analysis could be seen as providing a limit on the GWB amplitude, it depends on the timing value of these parameters, which is not fully independent of the simulated values. Therefore, we require an independent estimate of a timing parameter, as can be provided by VLBI, for example.

{ We have shown that many astrometric parameters are affected by a GWB. However, when comparing VLBI values \citep{dvt+08,cbv+09} to timing values \citep{vbs+08,vbc+09}, it is readily found that, with the exception of parallax, all VLBI parameter estimates are inconsistent with timing values at the many sigma level because of  calibrator uncertainty in the VLBI method.} With the parallax method we repeat the procedure described in the previous section for the timing parallax, but instead of comparing against the unperturbed timing value and uncertainty, we use the VLBI parameter and parallax . To accomplish this, we define an offset parameter
\begin{equation}
\Delta=\frac{|\pi_{\rm VLBI}-\pi_{\rm T2}|}{\sigma_{\rm VLBI}+\sigma_{\rm T2}},
\end{equation}
where $\pi_{\rm VLBI}$ and $\pi_{\rm T2}$ are the VLBI and \textsc{Tempo2} simulation values of parallax, respectively and the denominator is the sum of the formal uncertainties on the parameters. { The values of the VLBI parallaxes used in this paper are $\pi=6.396\pm0.054$ mas \citep{dvt+08} and $\pi=0.95^{+0.06}_{-0.05}$ mas \citep{cbv+09} for  PSR J0437--4715 and PSR J1713+0747, respectively.  It is important to note that we compare the VLBI values with those published in the original timing models \citep{vbs+08,vbc+09}, where the reduced chi-squared was normalised through application of backend specific error-multiplication factors.} While previous work has shown that a stochastic GWB will affect astrometric parameters \citep{gep+97,ksg+99,jaf04} measured by  VLBI such as parallax, it is known that  the angular deflections expected are on the order of the characteristic strain amplitude \citep[see e.g.][]{bf10}. Because typical amplitudes are $\sim 10^{-15}$ these effects are far less than the typical uncertainties on these astrometric parameters. Thus, we can treat  VLBI measurements as unaffected by the stochastic GWB. To place a limit, we determine the GWB amplitude in which 95\% of the realisations result in offset values in excess of 2. The factor of 2 is to ensure a 5\% false negative rate. The results from this method are shown in Table \ref{tab:amp}. It is clear from the table that the upper limits obtained from PSR J0437--4715 are more stringent than those obtained from PSR J1713+0747. This is to be expected, because the background does not corrupt the values of parallax as much for PSR J1713+0747, as shown in previous sections. Though using this method produces upper limits that are $\sim$10 times less stringent than the best published limits \citep{jhs+06} it provides an independent confirmation and is still constraining.

\subsection{Effects of the GWB spectral break}
Recent work \citep{svc08} has shown that the standard power law given in Equation \ref{eq:strain} for the GW strain only holds  for frequencies $f<10^{-8}$ Hz. Simulations were used to numerically formulate Equation \ref{eq:broken} which describes the GW strain for frequencies $f>10^{-8}$ Hz. GW detection strategies using pulsar timing are poised to detect the stochastic background at the lowest possible frequencies ($\sim 10^{-9}$ Hz), which is limited by the length of the data set, and thus still in the single power-law domain.  The parameters $f_0, \gamma$, and $h_0$ in Equation \ref{eq:broken} depend on different scenarios for MBH formation and evolution and can be found in Table 1 of \citet{svc08}. For clarity of our results here, the four models used are VHM \citep{vhm03}, KBD \citep{kbd04}, BVRhf \citep{bvr06}, and VHMhopk \citep{ln06}. 

A \textsc{Tempo2} plugin was developed to test whether this broken power-law has any affect on pulsar timing parameters or our upper limit calculations described in the previous section. This plugin  simulates a GWB using the simple power law spectrum for frequencies below the cutoff frequency of $10^{-8}$ Hz and then switch to the broken power law spectrum for frequencies greater than this cutoff. Since the parameters are model dependent, the plugin allows the user to input which model to use thereby setting the values of $f_0$ and $\gamma$; however, $h_0$ was simply determined by requiring that Equations \ref{eq:strain} and \ref{eq:broken} are continuous at the cutoff frequency. This step is required because producing an upper limit requires that we vary the value of the GW strain amplitude $A$ and thereby $h_0$. The strain spectrum for the above mentioned MBH formation scenarios is plotted in Figure \ref{fig:gwstrain}. It is clearly visible from the figure that there is a significant deviation from the power law strain spectrum. It also can be seen that the choice of model can make a considerable difference in the strain spectrum. Figure \ref{fig:gwpoints} shows the strain of the individual sources used to create a stochastic background using the VHM model. Note the significant deviation from power law strain spectrum (dashed line). { It should be noted here that we are using the average GW strain spectrum. In reality the spectral break comes from the small number of sources at higher frequencies, leading to a more jagged strain spectrum at these higher frequencies. While this method is sufficient for the work presented here, more realistic simulations will be very valuable for future studies of the stochastic GWB. However, such simulations are beyond the scope of this paper.} The simulation was run on the PSR J0437--4715 and PSR J1713+0747 data sets using the  same method as described above for all four models. The results are summarised in Table \ref{tab:amp}. It is obvious that implementation of  this broken power law does not significantly change the resulting upper limits. { This is to be expected since the low-frequency GW power dominates the corruption of the timing parameters. In fact, the induced rms of a GWB at the frequencies corresponding to the spectral break ($\sim 10^{-8}$ Hz) is around 10 ns \citep{sv10}, which is a factor of $\sim 20$ below our white noise level in both pulsars. We only include these results to conclusively show that the GWB spectral break has no effect on pulsar timing parameters.}

\subsection{Effects of ultralow--frequency gravitational waves}
\label{sec:kop}
Previous work \citep{kop97} has shown analytically that the secular variations of orbital period $P_b$ and of the semimajor axis $a$ projected onto the line of sight ($x=a\sin\iota$), where $\iota$ is the inclination angle, can be used to set upper limits on the energy density $\Omega_{\rm gw}$ in the ultralow-frequency ($10^{-12}$--$10^{-9}$ Hz) regime. This work derives upper limits in terms of the variance on the parameters $\dot{P}_{\rm b}$ and $\dot{x}$ indicating that a stochastic GWB in the ultralow-frequency range would significantly corrupt these parameters. To see if this proposition is consistent with our simulations, we perform the same steps described in  Section \ref{errors} except now we constrain the simulations to only the ultralow-frequency waves. Figure \ref{fig:lf} show the results of these simulations. Although it is not shown in the figure, as in the normal frequency case there is no biasing of these parameter distributions.  As shown in Figure 7, the effects of ultra-low-frequency GWs is substantially smaller than that of low-frequency GWs. That notwithstanding, the data sets affected by ultra-low-frequency GWs were more severely affected by phase-wrapping, but this is primarily because the simulated frequency range ($f < 10^{-9}$) implies that to first order, the GWB inserts a strong linear slope into the timing residuals, which can easily cause phase wraps (i.e. timing residuals beyond one pulse period). The low-frequency background ($f > 10^{-9}$) however, introduces timing residuals that are more akin to a random walk and while these typically have a strong linear component as well, it is  far less steep and therefore less likely to cause phase wrapping. The analysis on these two pulsars suggests that this method of placing upper limits on the GWB using the variances of $\dot{P}_{\rm b}$ and $\dot{x}$, respectively, is ineffective, or at best, far less constraining than simply using the variances in the low-frequency range.

\section{Summary}
\label{discussion}
We have introduced a  way to investigate the effects of { low-frequency noise in the form of} a stochastic background of GWs on pulsar timing parameters. { While this work focuses particularly on the stochastic GWB, the methods for investigating the noise/GW induced corruption of pulsar timing parameters could apply to any low-frequency noise source.} This method involves a Monte-Carlo simulation over different possible GWBs, adding them into the pulsar timing residuals, fitting for pulsar timing parameters, and producing distributions of various GW-affected pulsar parameters. These parameter distributions are then analysed to determine the overall effect that the GWB has on pulsar timing. { To summarise our results, we have:
\begin{itemize}
\item Replicated the \citet{vbs+08} results of red noise leakage and how it affects parameter uncertainties, showing that some pulsar parameters could be underestimated by up to a factor of $\sim 10$ for a GWB with an amplitude of $A=5\times 10^{-15}$.
\item Expanded on the above work with another pulsar (PSR J1713+0747), clarifying the importance of ecliptic latitude and the reliability of astrometric parameter measurements.
\item Shown that one can place upper limits on the stochastic GWB by comparing VLBI derived astrometric parameters to those obtained through pulsar timing.
\item Demonstrated practically that the spectral break (or its monochromatic components) are unlikely to affect pulsar timing before the SKA era.
\item Demonstrated that ultra-low frequency GWs have no obvious effect on pulsar timing parameters.
\item Demonstrated that even red noise at very low levels can leak into higher frequency terms and cause parameter corruptions. This lends a partial explanation for EFACs (a multiplicative error factor that is used to normalise the reduced chi--squared in the fitting process), even in a seemingly white data set.
\end{itemize}}
 Future work could investigate these timing effects on a larger sample of pulsars, including normal (non-MSP) pulsars to check for corruption of pulsar parameters though these effects should be smaller due to larger errors on the timing parameters.

\section*{Acknowledgments}
The Parkes telescope is part of the Australia Telescope which is funded by the Commonwealth of Australia for operation as a National Facility managed by CSIRO. JE is supported by the NSF PIRE program award No. 0968296. MAM is supported by the West Virginia EPSCoR program, Research Corporation for Scientific Advancement and the NSF PIRE program. JPWV is supported by the European Union under Marie Curie Intra-European Fellowship 236394. We thank the Parkes Pulsar Timing Array team for observing these pulsars and sharing these data as part of the IPTA agreement.

\bsp

\bibliographystyle{mnras}
\bibliography{apjjabb,bib}

\begin{thebibliography}{51}
\expandafter\ifx\csname natexlab\endcsname\relax\def\natexlab#1{#1}\fi

\bibitem[{Anholm} et~al.(2009){Anholm}, {Ballmer}, {Creighton}, {Price} \&
  {Siemens}]{abc+08}
{Anholm} M., {Ballmer} S., {Creighton} J.~D.~E., {Price} L.~R., {Siemens} X.,
  2009, Phys. Rev. D, 79, 8, 084030

\bibitem[{Begelman} et~al.(2006){Begelman}, {Volonteri} \& {Rees}]{bvr06}
{Begelman} M.~C., {Volonteri} M., {Rees} M.~J., 2006, {MNRAS}, 370, 289

\bibitem[{Book} \& {Flanagan}(2010)]{bf10}
{Book} L.~G., {Flanagan} {\'E}.~{\'E}., 2010, ArXiv e-prints

\bibitem[{Chatterjee} et~al.(2009){Chatterjee}, {Brisken}, {Vlemmings}
  et~al.]{cbv+09}
{Chatterjee} S., {Brisken} W.~F., {Vlemmings} W.~H.~T., et~al., 2009, ApJ, 698,
  250

\bibitem[{Coles} et~al.(2011){Coles}, {Hobbs}, Champion, Manchester \&
  Verbiest]{chc+10}
{Coles} W., {Hobbs} G., Champion D., Manchester R., Verbiest J., 2011,
  Submitted to \textit{MNRAS}

\bibitem[{Damour} \& {Vilenkin}(2001)]{dv01}
{Damour} T., {Vilenkin} A., 2001, Phys. Rev. D, 64, 6, 064008

\bibitem[{Deller} et~al.(2008){Deller}, {Verbiest}, {Tingay} \&
  {Bailes}]{dvt+08}
{Deller} A.~T., {Verbiest} J.~P.~W., {Tingay} S.~J., {Bailes} M., 2008, ApJL,
  685, L67

\bibitem[{Detweiler}(1979)]{det79}
{Detweiler} S., 1979, ApJ, 234, 1100

\bibitem[{Edwards} et~al.(2006){Edwards}, {Hobbs} \& {Manchester}]{ehm06}
{Edwards} R.~T., {Hobbs} G.~B., {Manchester} R.~N., 2006, {MNRAS}, 372, 1549

\bibitem[{Enoki} et~al.(2004){Enoki}, {Inoue}, {Nagashima} \&
  {Sugiyama}]{ein+04}
{Enoki} M., {Inoue} K.~T., {Nagashima} M., {Sugiyama} N., 2004, ApJ, 615, 19

\bibitem[{Foster} \& {Backer}(1990)]{fb90}
{Foster} R.~S., {Backer} D.~C., 1990, ApJ, 361, 300

\bibitem[{Gwinn} et~al.(1997){Gwinn}, {Eubanks}, {Pyne}, {Birkinshaw} \&
  {Matsakis}]{gep+97}
{Gwinn} C.~R., {Eubanks} T.~M., {Pyne} T., {Birkinshaw} M., {Matsakis} D.~N.,
  1997, ApJ, 485, 87

\bibitem[{Hellings} \& {Downs}(1983)]{hd83}
{Hellings} R.~W., {Downs} G.~S., 1983, ApJL, 265, L39

\bibitem[{Hobbs} et~al.(2010){Hobbs}, {Archibald}, {Arzoumanian}
  et~al.]{haa+10}
{Hobbs} G., {Archibald} A., {Arzoumanian} Z., et~al., 2010, Classical and
  Quantum Gravity, 27, 8, 084013

\bibitem[{Hobbs} et~al.(2009){Hobbs}, {Jenet}, {Lee} et~al.]{hjl+09}
{Hobbs} G., {Jenet} F., {Lee} K.~J., et~al., 2009, {MNRAS}, 394, 1945

\bibitem[{Hobbs} et~al.(2006){Hobbs}, {Edwards} \& {Manchester}]{hem06}
{Hobbs} G.~B., {Edwards} R.~T., {Manchester} R.~N., 2006, {MNRAS}, 369, 655

\bibitem[{Hotan} et~al.(2005){Hotan}, {Bailes} \& {Ord}]{hbo05}
{Hotan} A.~W., {Bailes} M., {Ord} S.~M., 2005, {MNRAS}, 362, 1267

\bibitem[{Jaffe}(2004)]{jaf04}
{Jaffe} A.~H., 2004, New Astronomy Reviews, 48, 1483

\bibitem[{Jaffe} \& {Backer}(2003)]{jb03}
{Jaffe} A.~H., {Backer} D.~C., 2003, ApJ, 583, 616

\bibitem[{Janssen} et~al.(2008){Janssen}, {Stappers}, {Kramer}, {Purver},
  {Jessner} \& {Cognard}]{jsk+08}
{Janssen} G.~H., {Stappers} B.~W., {Kramer} M., {Purver} M., {Jessner} A.,
  {Cognard} I., 2008, in { 40 Years of Pulsars: Millisecond Pulsars, Magnetars
  and More\/}, edited by {C.~Bassa, Z.~Wang, A.~Cumming, \& V.~M.~Kaspi}, vol.
  983 of { American Institute of Physics Conference Series\/},  633--635

\bibitem[{Jenet} et~al.(2009){Jenet}, {Finn}, {Lazio} et~al.]{jfl+09}
{Jenet} F., {Finn} L.~S., {Lazio} J., et~al., 2009, ArXiv e-prints

\bibitem[{Jenet} et~al.(2005){Jenet}, {Hobbs}, {Lee} \& {Manchester}]{jhl+05}
{Jenet} F.~A., {Hobbs} G.~B., {Lee} K.~J., {Manchester} R.~N., 2005, ApJL, 625,
  L123

\bibitem[{Jenet} et~al.(2006){Jenet}, {Hobbs}, {van Straten} et~al.]{jhs+06}
{Jenet} F.~A., {Hobbs} G.~B., {van Straten} W., et~al., 2006, ApJ, 653, 1571

\bibitem[{Jenet} et~al.(2004){Jenet}, {Lommen}, {Larson} \& {Wen}]{jll+04}
{Jenet} F.~A., {Lommen} A., {Larson} S.~L., {Wen} L., 2004, ApJ, 606, 799

\bibitem[{Kopeikin}(1997)]{kop97}
{Kopeikin} S.~M., 1997, Phys. Rev. D, 56, 4455

\bibitem[{Kopeikin} et~al.(1999){Kopeikin}, {Sch{\"a}fer}, {Gwinn} \&
  {Eubanks}]{ksg+99}
{Kopeikin} S.~M., {Sch{\"a}fer} G., {Gwinn} C.~R., {Eubanks} T.~M., 1999, Phys.
  Rev. D, 59, 8, 084023

\bibitem[{Koushiappas} et~al.(2004){Koushiappas}, {Bullock} \& {Dekel}]{kbd04}
{Koushiappas} S.~M., {Bullock} J.~S., {Dekel} A., 2004, {MNRAS}, 354, 292

\bibitem[{Leblond} et~al.(2009){Leblond}, {Shlaer} \& {Siemens}]{lss09}
{Leblond} L., {Shlaer} B., {Siemens} X., 2009, Phys. Rev. D, 79, 12, 123519

\bibitem[{Lodato} \& {Natarajan}(2006)]{ln06}
{Lodato} G., {Natarajan} P., 2006, {MNRAS}, 371, 1813

\bibitem[Lorimer \& Kramer(2005)]{handbook}
Lorimer D., Kramer M., 2005, Handbook of Pulsar Astronomy, vol.~4 of {
  Cambridge Observing Handbooks for Research Astronomers\/}, Cambridge
  University Press, Cambridge, U.K.; New York, U.S.A, 1st edn.

\bibitem[{Lorimer}(2008)]{l08}
{Lorimer} D.~R., 2008, Living Reviews in Relativity, 11, 8

\bibitem[{Manchester}(2008)]{m08}
{Manchester} R.~N., 2008, in { 40 Years of Pulsars: Millisecond Pulsars,
  Magnetars and More\/}, edited by {C.~Bassa, Z.~Wang, A.~Cumming, \&
  V.~M.~Kaspi}, vol. 983 of { American Institute of Physics Conference
  Series\/},  584--592

\bibitem[{Mignard} \& {Klioner}(2010)]{mk10}
{Mignard} F., {Klioner} S.~A., 2010, in { IAU Symposium\/}, edited by
  {S.~A.~Klioner, P.~K.~Seidelmann, \& M.~H.~Soffel}, vol. 261 of { IAU
  Symposium\/},  306--314

\bibitem[Press et~al.(1992)Press, Teukolsky, Vetterling \& Flannery]{pvt+92}
Press W.~H., Teukolsky S.~A., Vetterling W.~T., Flannery B.~P., 1992, Numerical
  recipes in C (2nd ed.): the art of scientific computing, Cambridge University
  Press, New York, NY, USA

\bibitem[{Romani}(1989)]{r89}
{Romani} R.~W., 1989, in { Timing Neutron Stars\/}, edited by {H.~{\"O}gelman
  \& E.~P.~J.~van den Heuvel},  113--+

\bibitem[{Sazhin}(1978)]{saz78}
{Sazhin} M.~V., 1978, {Soviet~Ast.}, 22, 36

\bibitem[{Sesana} \& {Vecchio}(2010)]{sv10}
{Sesana} A., {Vecchio} A., 2010, Classical and Quantum Gravity, 27, 8, 084016

\bibitem[{Sesana} et~al.(2008){Sesana}, {Vecchio} \& {Colacino}]{svc08}
{Sesana} A., {Vecchio} A., {Colacino} C.~N., 2008, {MNRAS}, 390, 192

\bibitem[{Shannon} \& {Cordes}(2010)]{sc10}
{Shannon} R.~M., {Cordes} J.~M., 2010, ApJ, 725, 1607

\bibitem[{Siemens} et~al.(2007){Siemens}, {Mandic} \& {Creighton}]{smc07}
{Siemens} X., {Mandic} V., {Creighton} J., 2007, Physical Review Letters, 98,
  11, 111101

\bibitem[{Splaver} et~al.(2005){Splaver}, {Nice}, {Stairs}, {Lommen} \&
  {Backer}]{sns+05}
{Splaver} E.~M., {Nice} D.~J., {Stairs} I.~H., {Lommen} A.~N., {Backer} D.~C.,
  2005, ApJ, 620, 405

\bibitem[{Standish}(2004)]{sta04}
{Standish} E.~M., 2004, A\&A, 417, 1165

\bibitem[{Stinebring} et~al.(1990){Stinebring}, {Ryba}, {Taylor} \&
  {Romani}]{srt+90}
{Stinebring} D.~R., {Ryba} M.~F., {Taylor} J.~H., {Romani} R.~W., 1990,
  Physical Review Letters, 65, 285

\bibitem[{Taylor} \& {Weisberg}(1982)]{tw82}
{Taylor} J.~H., {Weisberg} J.~M., 1982, ApJ, 253, 908

\bibitem[{van Haasteren} et~al.(2011){van Haasteren}, {Levin}, {Janssen}
  et~al.]{hlj+11}
{van Haasteren} R., {Levin} Y., {Janssen} G.~H., et~al., 2011, ArXiv e-prints

\bibitem[{van Haasteren} et~al.(2009){van Haasteren}, {Levin}, {McDonald} \&
  {Lu}]{vlm+09}
{van Haasteren} R., {Levin} Y., {McDonald} P., {Lu} T., 2009, {MNRAS}, 395,
  1005

\bibitem[{Verbiest} et~al.(2009){Verbiest}, {Bailes}, {Coles} et~al.]{vbc+09}
{Verbiest} J.~P.~W., {Bailes} M., {Coles} W.~A., et~al., 2009, {MNRAS}, 400,
  951

\bibitem[{Verbiest} et~al.(2008){Verbiest}, {Bailes}, {van Straten}
  et~al.]{vbs+08}
{Verbiest} J.~P.~W., {Bailes} M., {van Straten} W., et~al., 2008, ApJ, 679, 675

\bibitem[{Volonteri} et~al.(2003){Volonteri}, {Haardt} \& {Madau}]{vhm03}
{Volonteri} M., {Haardt} F., {Madau} P., 2003, ApJ, 582, 559

\bibitem[{Wen} et~al.(2011){Wen}, {Jenet}, {Yardley}, {Hobbs} \&
  {Manchester}]{wjy+11}
{Wen} Z.~L., {Jenet} F.~A., {Yardley} D., {Hobbs} G.~B., {Manchester} R.~N.,
  2011, ApJ, 730, 29

\bibitem[{Wyithe} \& {Loeb}(2003)]{wl03}
{Wyithe} J.~S.~B., {Loeb} A., 2003, ApJ, 590, 691

\end{thebibliography}

\newpage
\begin{table*}
\caption{PSR J0437--4715 Timing Model Parameters}
\label{tab:0437}
\centering
\begin{tabular}{p{.4\textwidth}cccc}
\hline\hline
Parameter Name and Units & Parameter Value & \textsc{Tempo2} Error$^{a}$ & GWB Induced Error$^{b}$ & Error Ratio\\
\hline
\multicolumn{5}{|c|}{Fit and Data Set}\\
\hline
MJD range \dotfill & 50191.0--53819.2 & & &\\
Number of TOAs \dotfill & 2847 & & &\\
Observation length, $T_{\rm obs}$ (yr)\dotfill & 9.9 & & &\\
rms timing residual ($\mu$s) \dotfill & 0.199 & & &\\
\hline
\multicolumn{5}{|c|}{Measured Quantities}\\
\hline  
Right ascension, $\alpha$ (J2000.0) \dotfill & 04\,37\,15.8147635 & 3 & 57 &  19\\
Declination, $\delta$ (J2000.0) \dotfill & -47\,15\,08.624170 & 3 & 60 & 20\\
Proper motion in $\alpha$, $\mu_{\alpha}$ (mas yr$^{-1}$) \dotfill & 121.453 & 1 & 13 & 13\\
Proper motion in $\delta$, $\mu_{\delta}$ (mas yr$^{-1}$) \dotfill & -71.457 & 1 & 13 & 13\\
Annual parallax, $\pi$ (mas) \dotfill & 6.65 & 7 & 98 & 14\\
Dispersion measure, DM (cm$^{-3}$\, pc) \dotfill & 2.6443 & 4 & 24 & 6\\
Pulse frequency, $\nu$ (ms) \dotfill & 173.68794618476804  & 3 & 2910 & 970\\
Pulse frequency derivative, $\dot{\nu}$ ($10^{-15}$\, s$^{2}$) \dotfill & -1.7284079 & 3 & 585 & 195\\
Orbital period, $P_{\rm b}$ (days) \dotfill & 5.74104646 & 108 & 216 & 2\\
Orbital period derivative, $\dot{P_{\rm b}}$, ($10^{-12}$) \dotfill & 3.73 & 2 & 6 & 3\\
Epoch of periastron passage, $T_{0}$ (MJD) \dotfill & 52009.852429 & 582 & 582 & 1\\
Projected semi-major axis, $x$ (s) \dotfill & 3.36669708 & 11 & 11 & 1\\
$\dot{x}$ ($10^{-14}$) \dotfill & 1.2$^{c}$ & 3$^{c}$ & $27^{c}$ & $9^{c}$\\
Longitude of periastron, $\omega_{0}$ (deg) \dotfill & 1.2224 & 365 & 365 & 1\\
Periastron advance, $\dot{\omega}$ (deg yr$^{-1}$) \dotfill & 0.01600 & 430 & 862 & 2\\
Longitude of ascension, $\Omega$ (deg) \dotfill & 207.8 & 23 & 92 & 4\\
Orbital inclination, $i$ (deg) \dotfill & 137.58 & 6 & 24 & 4\\
Companion mass, $m_{2}$ ($M_{\odot}$) \dotfill & 0.25 & 1 & 1 & 1\\
Orbital eccentricity, $e$ ($10^{-5}$) \dotfill & 1.9179 & 3 & 9 & 3\\
\hline
\multicolumn{5}{|c|}{Set Quantities}\\
\hline
Reference epoch for $P$, $\alpha$,  and $\delta$ determination (MJD) \dotfill & 52005 & & &\\
Reference epoch for DM determination (MJD) \dotfill & 53211 & & &\\
\hline
\end{tabular}
\raggedright{\footnotesize{\, $^{a}$\, Given uncertainties are 1 $\sigma$ values in the last digits of the parameter values.\\
\, $^{b}$  1 $\sigma$ uncertainties for a simulated GWB with amplitude $A=1\times 10^{-14}$.\\
\, $^{c}$\, Not part of original timing model in \citet{vbc+09}.  The full timing model including $\dot{x}$ was fitted in a separate simulation.}}
\end{table*}

\newpage
\begin{table*}
\caption{PSR J1713+0747 Timing Model Parameters}
\label{tab:1713}
\centering
\begin{tabular}{p{.4\textwidth}cccc}
\hline\hline
Parameter Name and Units & Parameter Value & \textsc{Tempo2} Error$^{a}$ & GWB Induced Error$^{b}$ & Error Ratio\\
\hline
\multicolumn{5}{|c|}{Fit and Data Set}\\
\hline
MJD range \dotfill & 49421.9--54546.8 & & &\\
Number of TOAs \dotfill & 392 & & &\\
Observation length, $T_{\rm obs}$ (yr)\dotfill & 14.0 & & &\\
rms timing residual ($\mu$s) \dotfill & 0.198 & & &\\
\hline
\multicolumn{5}{|c|}{Measured Quantities}\\
\hline  
Right ascension, $\alpha$ (J2000.0) \dotfill & 17\,13\,49.532628 & 1 &  6 & 6 \\
Declination, $\delta$ (J2000.0) \dotfill & +07\,47\,37.50165 & 3 & 12 &  4\\
Proper motion in $\alpha$, $\mu_{\alpha}$ (mas yr$^{-1}$) \dotfill & 4.924 & 5 & 35 & 7\\
Proper motion in $\delta$, $\mu_{\delta}$ (mas yr$^{-1}$) \dotfill & -3.82& 1 & 8 & 8\\
Annual parallax, $\pi$ (mas) \dotfill & 0.94 & 5 & 15 & 3\\
Dispersion measure, DM (cm$^{-3}$\, pc) \dotfill & 15.9915 & 1 & 3 & 3\\
Pulse frequency $\nu$ (Hz) \dotfill & 218.8118404414362 & 2 &  1786 &  893\\
Pulse frequency derivative, $\dot{\nu}$ ($10^{-16}$ s$^{2}$) \dotfill & -4.08379 & 2 &  134 &  67\\
Orbital period, $P_{\rm b}$ (days) \dotfill & 67.825130963 & 9 & 27 & 3\\
Orbital period derivative, $\dot{P_{\rm b}}$, ($10^{-13}$) \dotfill & 41 & 10 & 60 & 6\\
Epoch of periastron passage, $T_{0}$ (MJD) \dotfill & 54303.6328 & 4 & 8 & 2\\
Projected semi-major axis, $x$ (s) \dotfill & 32.3424236 & 2 & 4 & 2\\
$\dot{x}$ ($10^{-15}$) \dotfill & -1.6$^{c}$ & 5$^{c}$ & 35$^{c}$ & 7$^{c}$\\
Longitude of periastron, $\omega_{0}$ (deg) \dotfill & 176.190 & 2 & 12 & 6\\
Longitude of ascension, $\Omega$ (deg) \dotfill & 67 & 9 & 63 & 7\\
Orbital inclination, $i$ (deg) \dotfill & 78.6 & 9 & 72 & 8\\
Companion mass, $m_{2}$ ($M_{\odot}$) \dotfill & 0.20 & 2 & 12 & 6\\
Orbital eccentricity, $e$ ($10^{-5}$) \dotfill & 7.4940 & 3 &  18 & 6\\
\hline
\multicolumn{5}{|c|}{Set Quantities}\\
\hline
Reference epoch for $P$, $\alpha$,  and $\delta$ determination (MJD) \dotfill & 54312 & & &\\
Reference epoch for DM determination (MJD) \dotfill & 54312 & & &\\
\hline
\end{tabular}
\raggedright{\footnotesize{\, $^{a}$\, Given uncertainties are 1 $\sigma$ values in the last digits of the parameter values.\\
\, $^{b}$  1 $\sigma$ uncertainties for a simulated GWB with amplitude $A=1\times 10^{-14}$.\\
\, $^{c}$\, Not part of original timing model in \citet{vbc+09}.  The full timing model including $\dot{x}$ was fitted in a separate simulation.}}
\end{table*}

\newpage
\begin{table*} 
\caption{Upper limits on the stochastic GWB amplitude (measured in units of $10^{-13}$) at the 2--sigma level for different power law models for pulsars PSR J0437--4715 and PSR J1713+0747.}
\vspace{2mm}
\label{tab:amp}
\centering
\begin{tabular}{l|cc}
\hline\hline
Spectrum& PSR J0437--4715 & PSR J1713+0747 \\ 
\hline
$\alpha=-2/3$ & 0.91 & 1.3\\ 
VHM & 0.91 & 1.3\\
BVRhf & 0.91 & 1.2\\
VHMhopk & 0.92 & 1.2\\
KBD & 0.91 & 1.3\\
\hline
\end{tabular}
\end{table*}

\newpage
\begin{figure*}
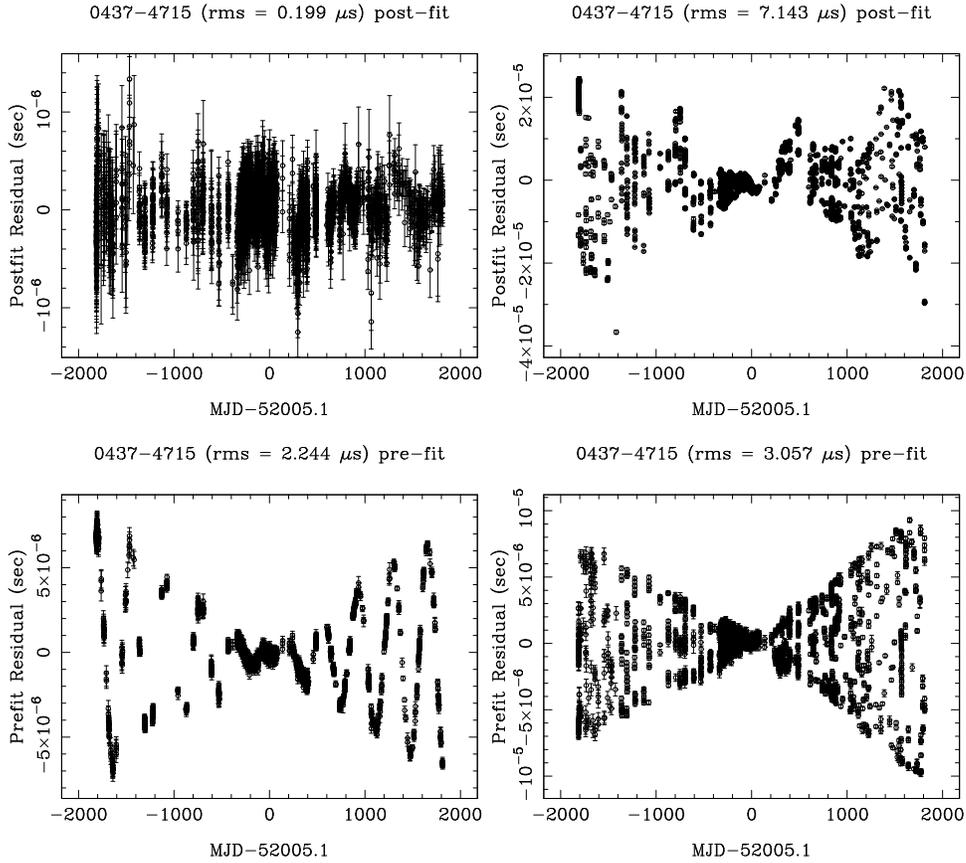

  \begin{center}
    \subfigure{\label{fig:fit}\includegraphics[scale=0.3,angle=270]{fig1a.ps}}
    \subfigure{\label{fig:gw}\includegraphics[scale=0.3,angle=270]{fig1b.ps}} \\
    \subfigure{\label{fig:pmpx}\includegraphics[scale=0.3,angle=270]{fig1c.ps}} 
    \subfigure{\label{fig:pbom}\includegraphics[scale=0.3,angle=270]{fig1d.ps}} 
  \end{center}
  \caption{\small{Plots of pulsar timing residuals of pulsar PSR J0437--4715 \citep{vbs+08} for different situations. Top left: no GWB added. We see here that the rms is low and the noise is relatively white. Top right: post-fit residuals after adding a GWB with amplitude $A=5\times 10^{-13}$.  Bottom left:  no GWB added but using the GW-induced values of parallax and proper motion in our timing model. This shows strong periodic structure at periods $\sim$ 1 yr. Bottom right: no GWB added but using GW-induced values of  $\dot{P}_{\rm b}$, $\omega$, and $\dot{\omega}$. }}
  \label{fig:fits}
\end{figure*}

\newpage
\begin{figure*}
\begin{center}
\includegraphics[scale=.75,angle=270]{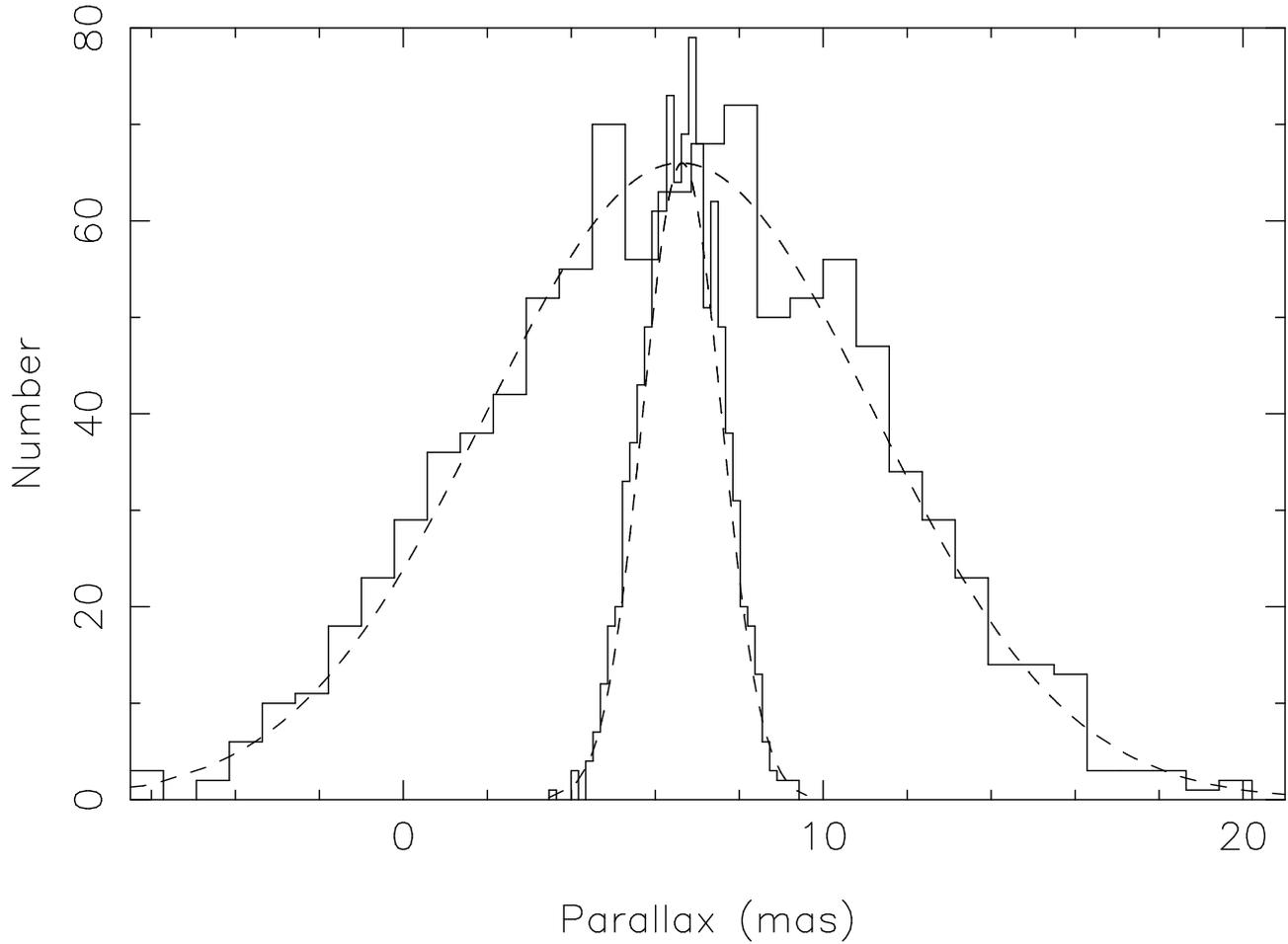}
\caption{\small{Distribution of the timing-derived parallax  for PSR J0437--4715 as a result of fitting a timing model to 1000 realisations of residuals with a stochastic GWB added. Plotted here are the histograms (solid lines) of the fitted parameters as well as  Gaussian curves (dashed lines) produced from the amplitude of the histogram, the mean of the distribution, and the standard deviation. It can be seen that a Gaussian fits the histogram very well. Here the histogram and Gaussian curve for the wider distribution is for a GWB amplitude of $A\sim 5\times 10^{-14}$ and the narrower distribution is for a GWB of $A\sim 1\times 10^{-14}$.}}
\label{fig:gauss}
\end{center}
\end{figure*}

\newpage
\begin{figure*}
\begin{center}
\includegraphics[scale=0.75,angle=270]{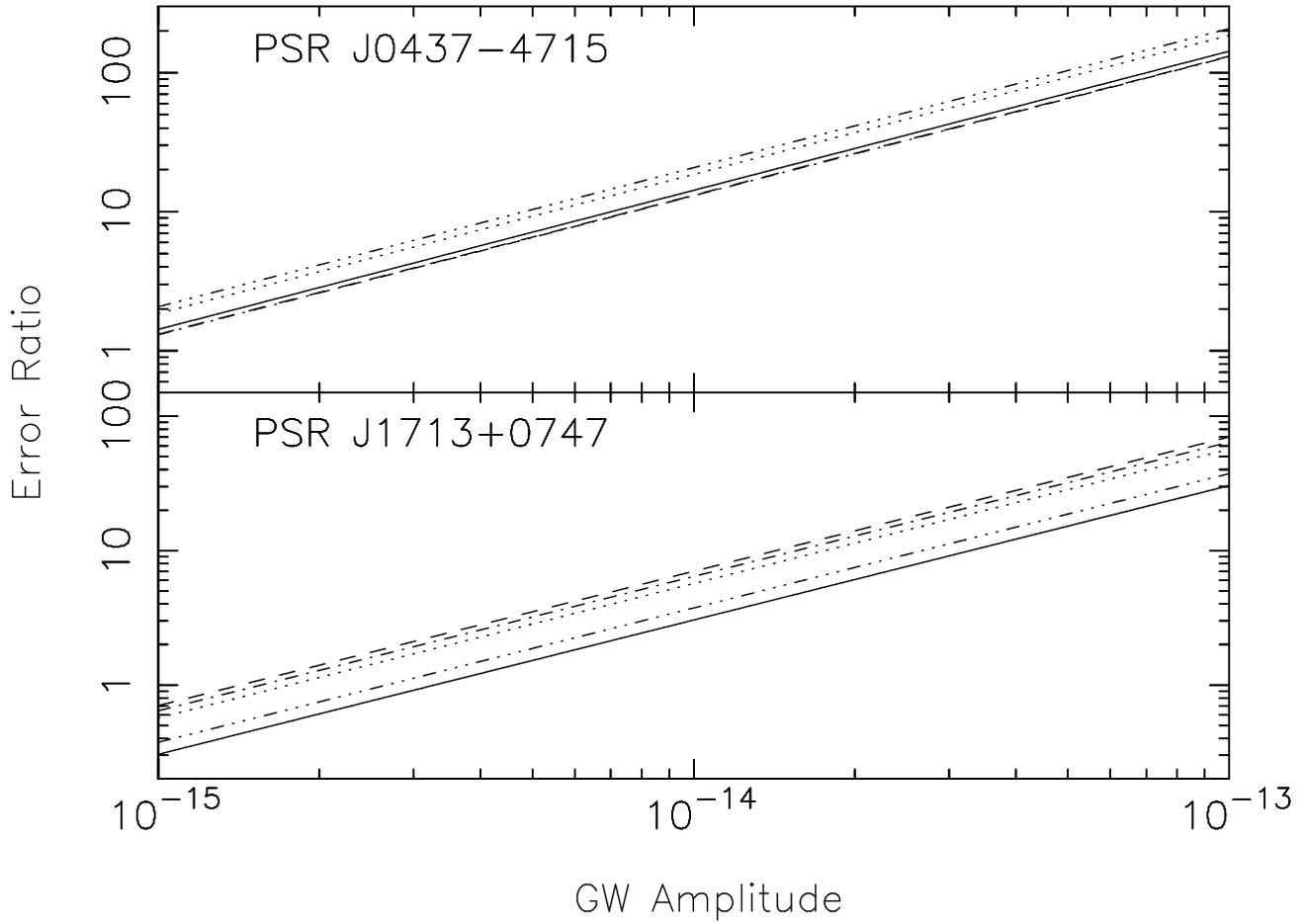}
\caption{\small{Plot of error ratio (ratio of standard deviation of distribution to the unperturbed error) vs. GWB amplitude { for { selected} timing model parameters, excluding spin period and spindown (see Tables \ref{tab:0437} and \ref{tab:1713})} of PSR J0437--4715 and PSR J1713+0747. (Solid line: parallax, Dashed line: proper motion in right ascension, Dot-dashed line: proper motion in declination, Dotted line:  right ascension, Dash-triple dot: declination.) It should be noted that the proper motion in right ascension and declination curves overlap for PSR J0437--4715.}}
\label{fig:err}
\end{center}
\end{figure*}

\newpage
\begin{figure*}
\begin{center}
\includegraphics[scale=0.75,angle=270]{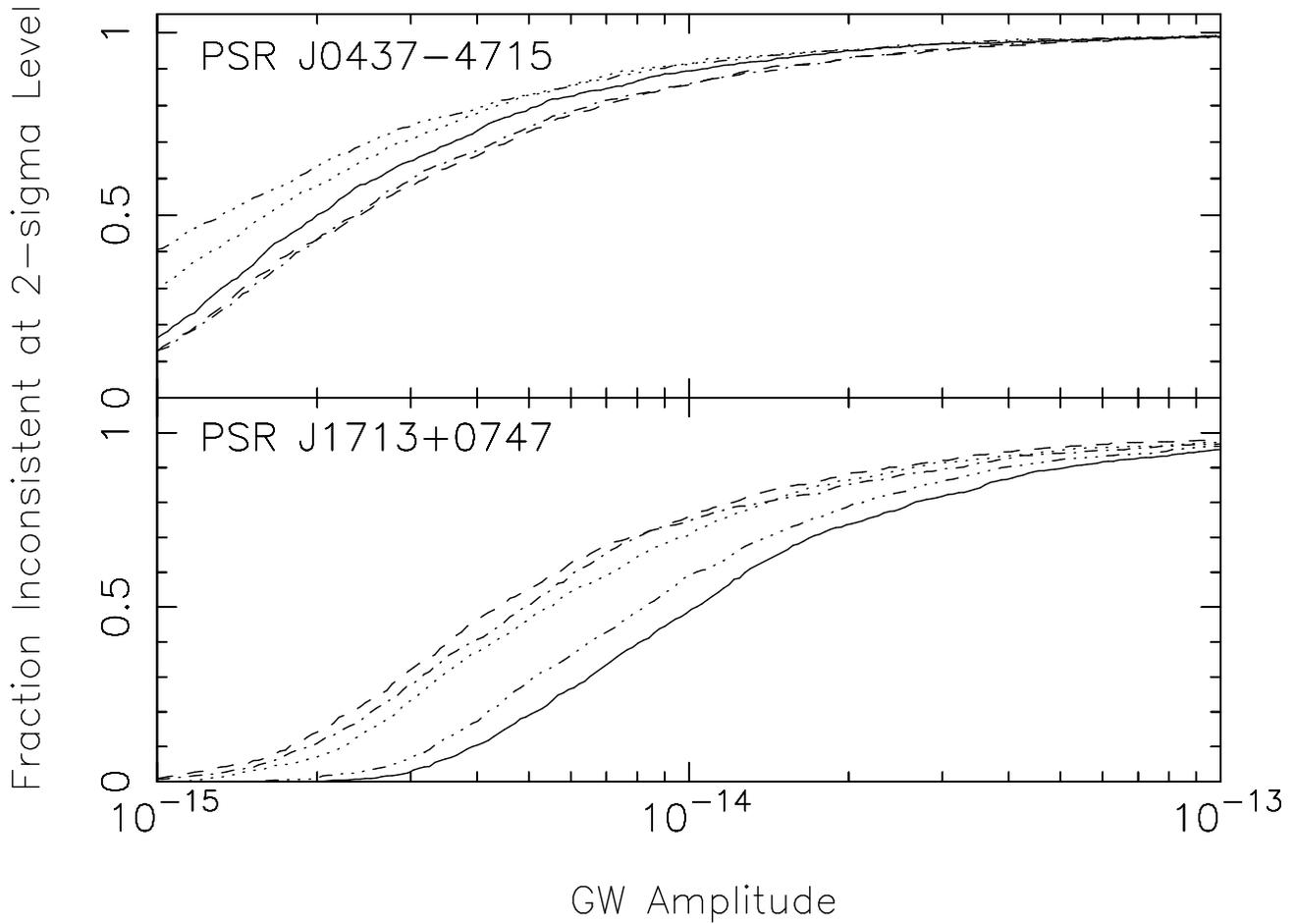}
\caption{\small{Plot of detection significance vs. GWB amplitude for PSRs PSR J0437--4715 and PSR J1713+0747 with parameters assumed to have the values listed in \citet{vbs+08,vbc+09}.  The method for obtaining these plots is described in the text. These plots represent detection rates with false negative rates of 5\% against GWB amplitude. Solid line: parallax, Dashed line: proper motion in right ascension, Dot-dashed line: proper motion in declination, Dotted line:  right ascension, Dash-triple dot: declination.)}}
\label{fig:detection}
\end{center}
\end{figure*}

\newpage
\begin{figure*}
\begin{center}
\includegraphics[scale=.75,angle=0]{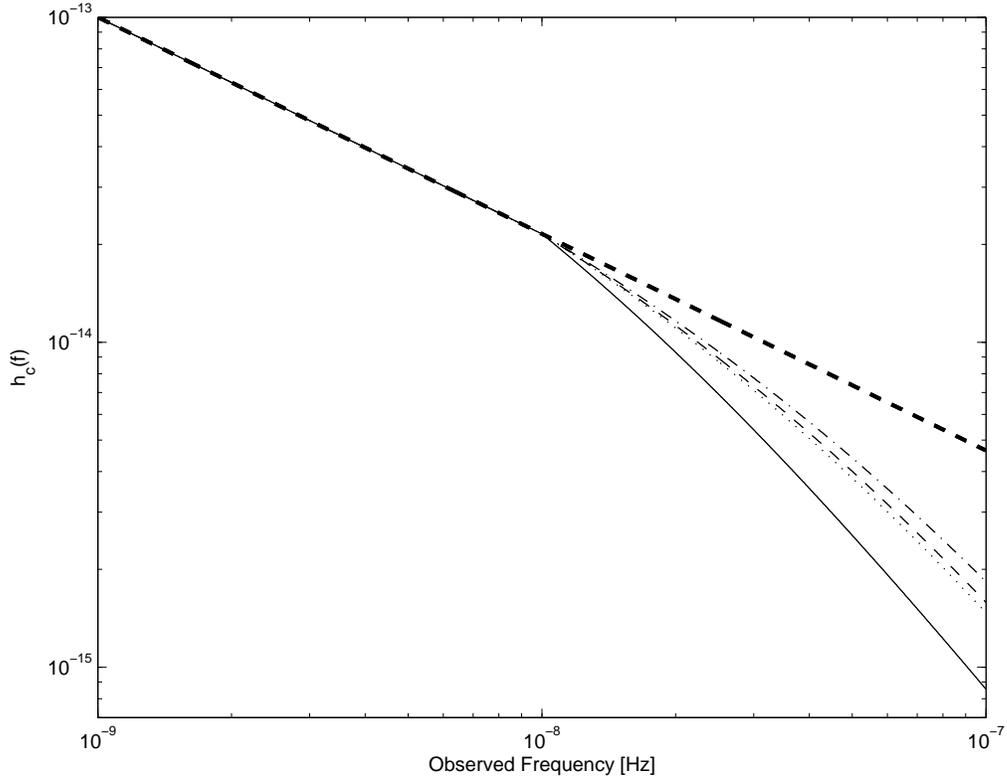}
\caption{\small{Plot of $h_c(f)$ against observed GW frequency. Here we show the naive power law spectrum from Equation \ref{eq:strain} and the broken power spectrum from Equation \ref{eq:broken} using different SMBH assembly models. (Thick dashed line: Normal power law spectrum, solid line: VHM model, dashed line: VHMhopk model, dash-dot line: KBD model, dotted line: BVRhf model.)}}
\label{fig:gwstrain}
\end{center}
\end{figure*}

\newpage
\begin{figure*}
\begin{center}
\includegraphics[scale=.5,angle=270]{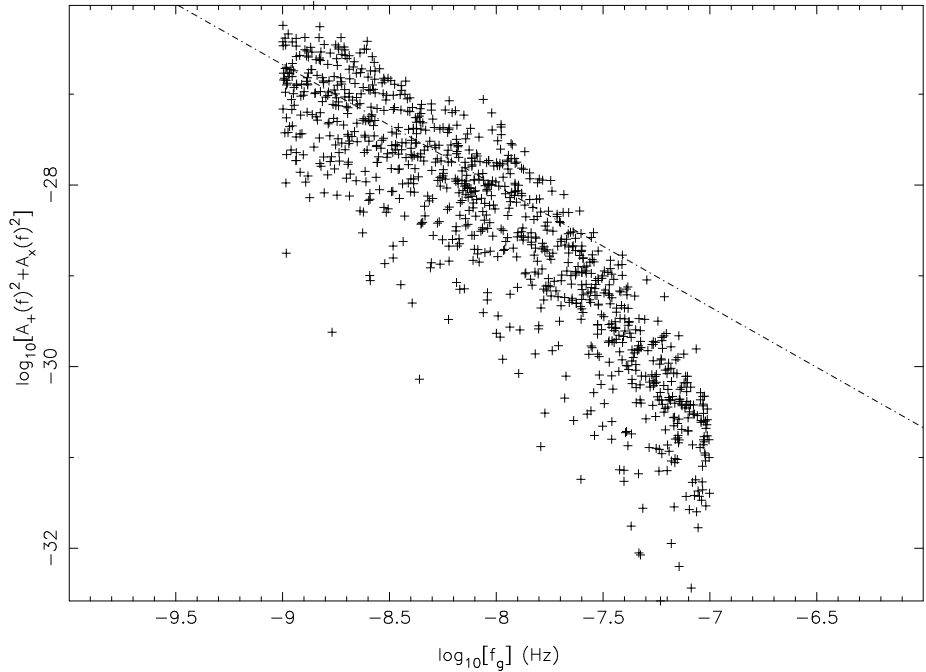}
\caption{\small{Plot of GW strain squared vs. GW frequency. The dotted line is the naive spectrum from Equation \ref{eq:strain} for the GW strain spectrum. The points are individual SMBH binary sources emitting at random frequencies in the range $10^{-9}$--$10^{-7}$ Hz generated by using Equation \ref{eq:broken} and the VHM model. Note that there is a significant deviation in the spectrum around $f\gtrsim10^{-8}$ Hz.}}
\label{fig:gwpoints}
\end{center}
\end{figure*}

\newpage
\begin{figure*}
\begin{center}
\subfigure{\label{fig:hilo0437}}\includegraphics[scale=.7,angle=0]{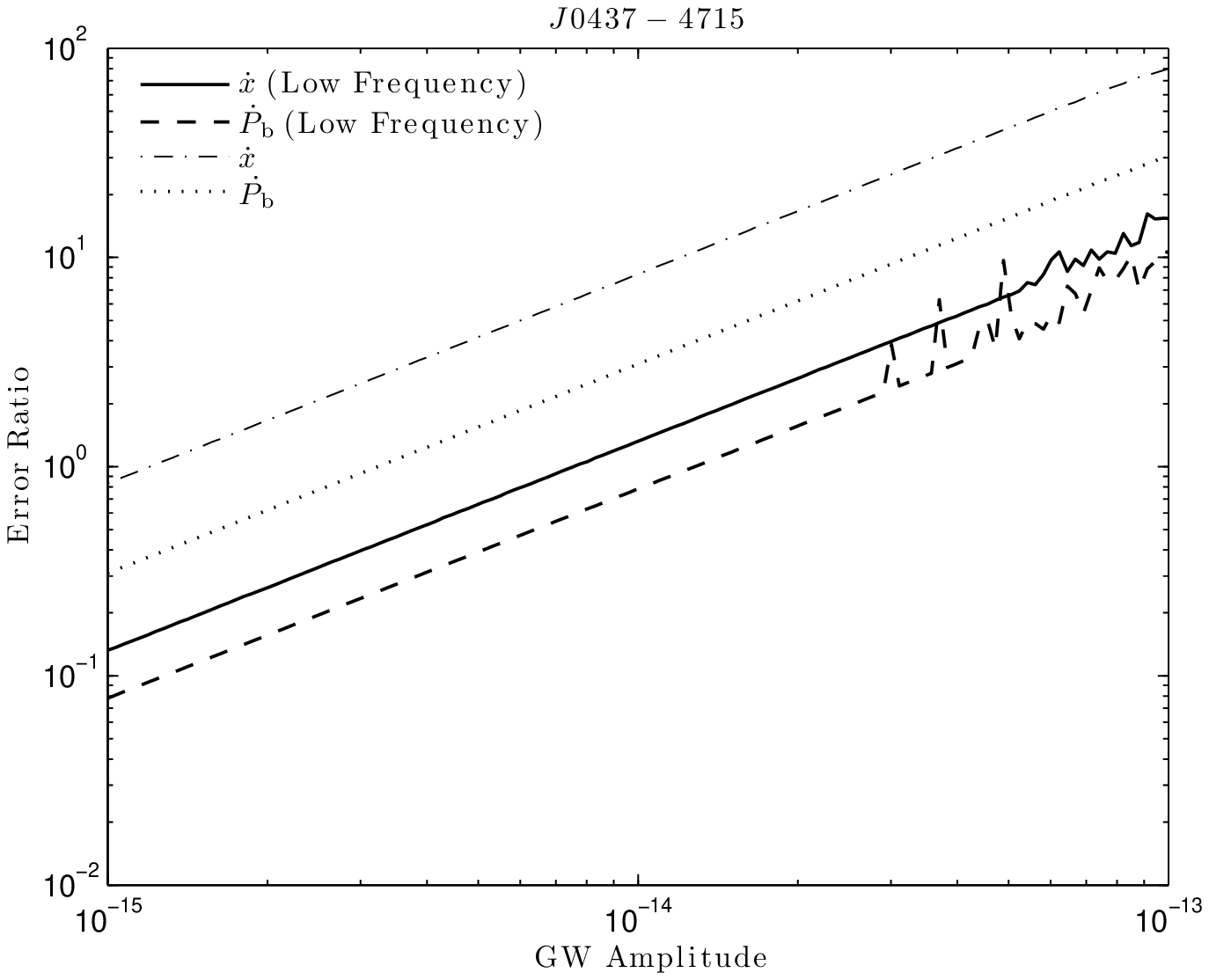}
\subfigure{\label{fig:hilo1713}}\includegraphics[scale=.7]{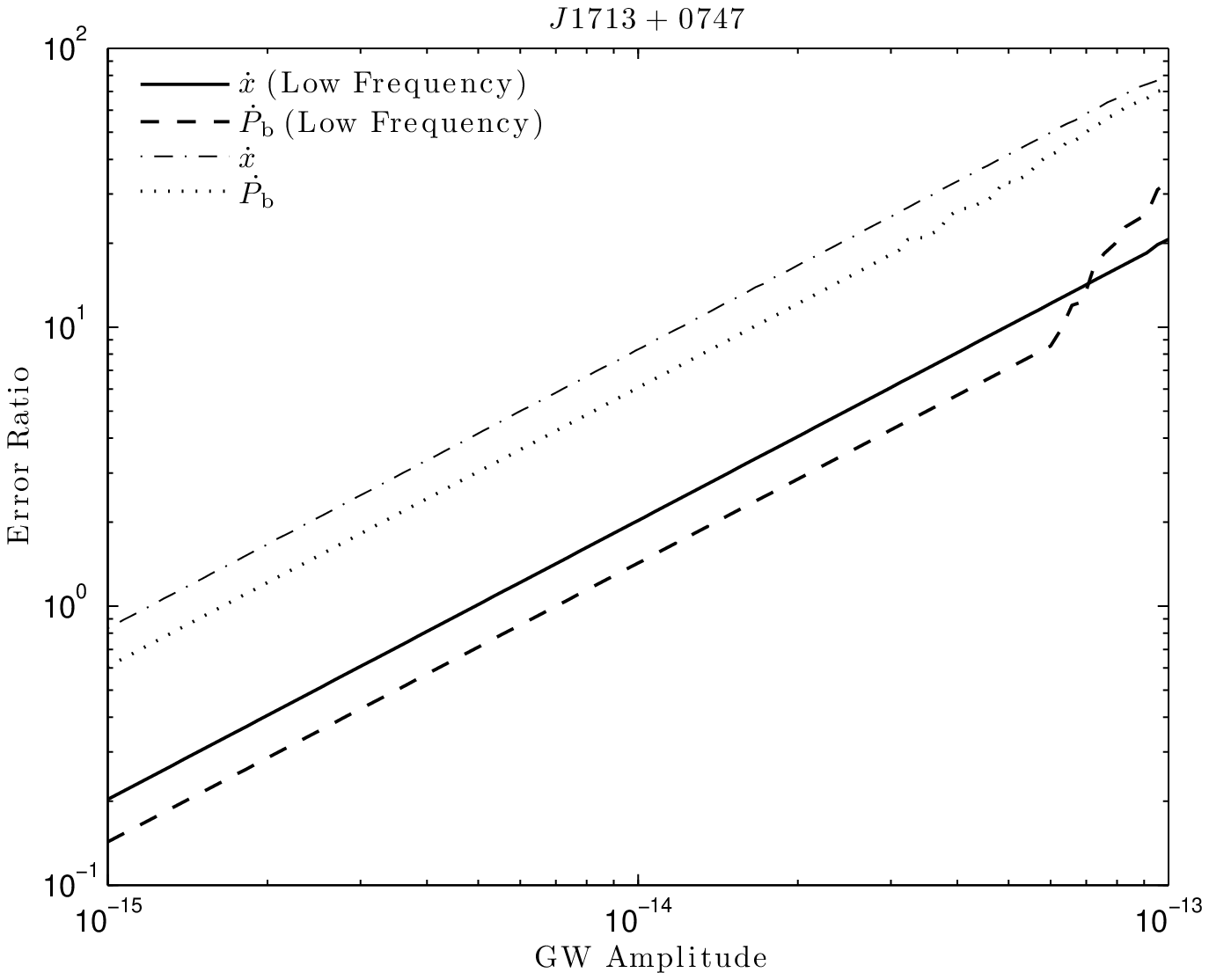}
\caption{\small{Plot of error ratio vs. GWB amplitude for $\dot{P}_{\rm b}$ and $\dot{x}$ in the normal ($10^{-9}$--$10^{-7}$ Hz) and ultralow ($10^{-12}$--$10^{-9}$ Hz) frequency ranges for  PSR J0437--4715 and PSR J1713+0747. The departure from a linear trend, seen at the highest simulated amplitudes, is caused by phase-wrapping that occurs when the simulated GW signature exceeds a few pulse periods. Such wrapping effectively randomises the results from the least-squares fit and produces outlier results in the Monte-Carlo iteration in which it occurs. This in turn distorts the statistics derived from that simulation and therefore corrupts the curves displayed here. We have used phase-tracking methods to avoid such wraps, but towards $A = 10^{-13}$ the effect becomes so large in certain Monte--Carlo iterations that phase-tracking is effectively rendered useless. }}
\label{fig:lf}
\end{center}
\end{figure*}

\label{lastpage}

\end{document}